\documentclass[12pt]{article}
\usepackage{graphicx}

\newcommand\pubnumber{UAB-FT-687}
\newcommand\pubdate{\today}

\def\UAB{Grup de F\'{\i}sica Te\`orica and IFAE,\\
                 Universitat Aut\`onoma de Barcelona,
                 E-08193 Bellaterra (Barcelona), SPAIN}
\def\ICREA{Instituci\'o Catalana de Recerca i Estudis Avan\c cats (ICREA),\\
                     IFAE and Grup de F\'{\i}sica Te\`orica,\\ 
                     Universitat Aut\`onoma de Barcelona, E-08193 Bellaterra (Barcelona), SPAIN}
\def\support{\footnote{Work supported by 
the Ministerio de Ciencia e Innovaci\'on under grant CICYT-FEDER-FPA2008-01430,
the EU Contract No.~MRTN-CT-2006-035482, ``FLAVIAnet'',
the Spanish Consolider-Ingenio 2010 Programme CPAN (CSD2007-00042), and 
the Generalitat de Catalunya under grant SGR2009-00894.}}

\def\Title#1{\begin{center} {\Large #1 } \end{center}}
\def\Author#1{\begin{center}{ \sc #1} \end{center}}
\def\Address#1{\begin{center}{ \it #1} \end{center}}

\newcommand\pubblock{\rightline{\begin{tabular}{l} \pubnumber\\
         \pubdate  \end{tabular}}}
\newenvironment{Abstract}{\begin{quotation}  }{\end{quotation}}
\newenvironment{Presented}{\begin{quotation} \begin{center} 
             PRESENTED AT\end{center}\bigskip 
      \begin{center}\begin{large}}{\end{large}\end{center} \end{quotation}}
\def\Acknowledgements{\bigskip  \bigskip \begin{center} \begin{large}
             \bf ACKNOWLEDGEMENTS \end{large}\end{center}}

\def\beq{\begin{equation}}
\def\eeq#1{\label{#1}\end{equation}}
\def\eeqn{\end{equation}}

\def\beqa{\begin{eqnarray}}
\def\eeqa#1{\label{#1}\end{eqnarray}}
\def\eeqan{\end{eqnarray}}

\let\bar=\overbar

\def\Dslash{\not{\hbox{\kern-4pt $D$}}}
\def\dslash{\not{\hbox{\kern-2pt $\del$}}}

\def\msb{{\bar{\ssstyle M \kern -1pt S}}}

\def\beq {\begin{equation}}
\def\eeq {\end{equation}}
\def\bea {\begin{eqnarray}}
\def\eea {\end{eqnarray}}

\def\nn {\nonumber}

\def\tauKpi{\tau\to K \pi\nu_\tau}

\begin{document}
\begin{titlepage}
\pubblock

\vfill
\Title{The $K\pi$ vector form factor and constraints from $K_{l3}$ decays\support}
\vfill
\Author{Diogo R.~Boito, Rafel Escribano\footnote{Speaker}}
\Address{\UAB}
\Author{Matthias Jamin}
\Address{\ICREA}
\vfill
\begin{Abstract}
The slope and curvature parameters of the $K\pi$ vector form factor, $F_+^{K\pi}$,
are fitted to the data on $\tauKpi$ and $K_{l3}$ decays yielding
$\lambda_+'=(25.49 \pm 0.31) \times 10^{-3}$ and $\lambda_+''= (12.22 \pm 0.14) \times 10^{-4}$.
The pole position of the $K^*(892)^\pm$ is found to be at
$m_{K^*(892)^\pm}= 892.0\pm 0.5$ MeV and $\Gamma_{K^*(892)^\pm}= 46.5 \pm1.1$ MeV.
The phase-space integrals relevant for $K_{l3}$ analyses and
the $P$-wave isospin-1/2 $K\pi$ phase-shift threshold parameters are also calculated.
\end{Abstract}
\vfill
\begin{Presented}
6th International Workshop on the CKM Unitarity Triangle,
University of Warwick, UK, 6-10 September 2010
\end{Presented}
\vfill
\end{titlepage}
\def\thefootnote{\fnsymbol{footnote}}
\setcounter{footnote}{0}

\section{Introduction}
The non-perturbative physics of $K\to\pi l\nu_l$ ($K_{l3}$) and $\tauKpi$ decays is governed by two Lorentz-invariant $K\pi$ form factors, namely the vector, denoted $F^{K\pi}_+(q^2)$,
and the scalar, $F^{K\pi}_0(q^2)$. 
A good knowledge of these form factors paves the way for the determination of many parameters of the Standard Model, such as the quark-mixing matrix element $|V_{us}|$ obtained from $K_{l3}$ decays \cite{LR84}, or the strange-quark mass $m_s$ determined from the scalar QCD strange spectral function \cite{JOP4}.

Until recently, the main source of experimental information on $K\pi$ form factors have been
$K_{l3}$ decays.
Lately, five experiments have collected data on
semileptonic and leptonic $K$ decays: BNL-E865, KLOE, KTeV, ISTRA+, and NA48.
Additional knowledge on the $K\pi$ form factors can be gained from the dominant
Cabibbo-suppressed $\tau$ decay: the channel $\tauKpi$.
A detailed spectrum for $\tau\to K_S\pi^-\nu_\tau$ produced and analyzed by Belle was published in 2007 \cite{Belle}.
Also, preliminary BaBar spectra with similar statistics have appeared recently in conference
proceedings \cite{Babar} and, finally, BESIII should produce results for this decay in the future
\cite{BESIII}.
The new data sets provide the substrate for up-to-date theoretical analyses of the $K\pi$ form
factors.
In Ref.~\cite{BEJ} we have performed a reanalysis of the $\tauKpi$ spectrum of \cite{Belle}.
More recently, we carried out an analysis with restrictions from $K_{l3}$ experiments
\cite{BEJ2010}.

On the theory side, the knowledge of these form factors consists of two tasks.
The first of them is to determine their value at the origin, $F_{+,0}(0)$,
crucial in order to disentangle the product $|V_{us}|F_{+,0}(0)$.
Historically, chiral perturbation theory has been the main tool to study $F_{+,0}(0)$,
but recently lattice QCD collaborations have produced more accurate results for this quantity
\cite{Lattice}.
Second, one must know the energy dependence of the form factors,
which is required when calculating phase-space integrals for $K_{l3}$ decays or when
analyzing the detailed shape of the $\tauKpi$ spectrum.
In our work we concentrate on the latter aspect of the problem and therefore it is convenient to
introduce form factors normalized to one at the origin\footnote{
From now on we refrain from writing the superscript $K\pi$ on the form factors.}
\beq
\tilde F_{+,0}(q^2)=F_{+,0}(q^2)/F_{+,0}(0)\ .
\eeq

A salient feature of the form factors in the kinematical region relevant for $K_{l3}$ decays,
{\emph i.e.}~$m_l^2<q^2<(m_K-m_\pi)^2$, is that they are real. 
Within the allowed phase-space they admit a Taylor expansion and the energy dependence is customarily translated into constants $\lambda_{+,0}^{(n)}$ defined as
\beq
\tilde F_{+,0}(q^2)=1+\lambda_{+,0}' \frac{q^2}{m_{\pi^-}^2}+
\frac{1}{2}\lambda_{+,0}'' \left( \frac{q^2}{m_{\pi^-}^2}\right)^2+\cdots\ .
\label{FFTaylor}
\eeq
In $\tauKpi$ decays, however, since $(m_K+m_\pi)^2<q^2<m_\tau^2$, one deals with a different kinematical regime in which the form factors develop imaginary parts, rendering the expansion of
Eq.~(\ref{FFTaylor}) inadmissible.
One must then resort to more sophisticated treatments.
Moreover, in order to fully benefit from the available experimental data,
it is desirable to employ representations of the form factors that are valid for both
$K_{l3}$ and $\tauKpi$ decays.
Dispersive representations of the form factors provide a powerful tool to achieve this goal.

From general principles, the form factors must satisfy a dispersion relation.
Supplementing this constraint with unitarity, the dispersion relation has a well-known closed-form solution within the elastic approximation referred to as the Omn\`es representation \cite{Omnes}. Although simple, this solution requires the detailed knowledge of the phase of $F_+(s)$ up to infinity, which is unrealistic.
An advantageous strategy to circumvent this problem is the use of additional  subtractions, as done, for instance, for the pion form factor in Ref.~\cite{PP2001}.
Subtractions in the dispersion relation entail a suppression of the integrand in the dispersion integral for higher energies. 
The outcome of these tests is that for our purposes an optimal description of $F_+(s)$ is reached with three subtractions and two resonances.
Here we quote the resulting  expression
\beq
\tilde F_+(s) = \exp\Bigg[
\alpha_1\frac{s}{m_{\pi^-}^2} +\frac{1}{2}\alpha_2\frac{s^2}{m_{\pi^-}^4}
+\frac{s^3}{\!\pi}\int\limits^{s_{\rm cut}}_{s_{K\pi}}ds'\frac{\delta(s')}{(s')^3(s'-s-i0)}\Bigg]\ .
\label{dispFF}
\eeq
In the last equation, $s_{K\pi}=(m_{K^0}+m_{\pi^-})^2$ and the two subtraction constants
$\alpha_1$ and $\alpha_2$ are related to the Taylor expansion of Eq.~(\ref{FFTaylor}) as
$\lambda_+'=\alpha_1$ and $\lambda_+''=\alpha_2+\alpha_1^2$.
It is opportune to treat them as free parameters that capture our ignorance of the higher energy part
of the integral.
The constants $\lambda_+'$ and $\lambda_+''$ can then be determined through the fit.
The main advantage of this procedure, advocated for example in
Refs.~\cite{BEJ,PP2001,Bernardetal},
is that the subtraction constants turn out to be less model dependent as they are determined by the best fit to the data.
It is important to stress that Eq.~(\ref{dispFF}) remains valid beyond the elastic approximation
provided $\delta(s)$ is the phase of the form factor, instead of the corresponding scattering phase. But, of course, in order to employ it in practice we must have a model for the phase.
As described in detail in Ref.~\cite{BEJ},
we take a form inspired by the RChT treatment of Refs.~\cite{JPP20062008} with two vector resonances.
For the detailed expressions we refer to the original works.
With Eq.~(\ref{dispFF}), the transition from the kinematical region of $\tauKpi$ to that of $K_{l3}$ decays is straightforward and the dominant low-energy behavior of $F_+(s)$ is encoded in
$\lambda_+'$ and $\lambda_+''$.
The cut-off $s_{\rm cut}$ in the dispersion integral is introduced to quantify the suppression of the higher energy part of the integrand. The stability of the results is checked varying this cut-off in a wide range from $1.8\, \mbox{GeV} < s_{\rm cut} < \infty$.

In $\tauKpi$ decays, the scalar form factor is suppressed kinematically.
Albeit marginal, the contribution from $F_0$ cannot be neglected in the lower energy part of the spectrum.
Here, we keep this contribution fixed using the results for $F_0$ from the coupled-channel
dispersive analysis of Refs.~\cite{JOP4,JOP1}.

\section{Fits to \mbox{\boldmath $\tauKpi$} with constraints from \mbox{\boldmath $K_{l3}$}}

The analysis of the spectrum for $\tauKpi$ produces a wealth of physical results, many of them with great accuracy, e.g., the mass and width of the $K^*(892)$.
We have advocated by means of Monte Carlo simulations that a joined analysis of $\tauKpi$ and $K_{l3}$ spectra further constrains the low-energy part of the vector form-factor yielding results with a better precision \cite{BEJ}.
This idea was pursued in our recent work \cite{BEJ2010}.

In order to include the experimental information available from $K_{l3}$ decays 
---and for the want of true unfolded data sets from these experiments--- 
we adopt the following strategy\footnote{
For a detailed discussion of the fit procedure we refer to \cite{BEJ2010}.}.
In our fits, the $\chi^2$ that is to be minimized contains a standard part from the $\tauKpi$ spectrum and a piece which constrains the parameters $\lambda_+^{(' ,'')}$ using information from $K_{l3}$
experiments.
For the latter experimental values we employ the results of the compilation of $K_L$ analyses performed by Antonelli {\it et al.}~for the FlaviaNet Working Group on Kaon Decays in Ref.~\cite{Antonelli10}:
$\lambda_+^{\prime\, \rm exp}=(24.9\pm1.1)\times 10^{-3}$, 
$\lambda_+^{\prime\prime\, \rm exp}=(16\pm 5)\times 10^{-4}$ and
$\rho_{\lambda_+^{\prime},\lambda_+^{\prime \prime}}=-0.95$.


\subsection{Results}
From the minimization of the $\chi^2$ a collection of physical results can be derived.
Some of them are obtained directly from the fit, such as $\lambda_+'$ and $\lambda_+''$
and the mass and width of the $K^*(892)$.
With the form factor under control,
one can then obtain other results such as the phase-space integrals for $K_{l3}$ decays.
Here, we present the main results of Ref.~\cite{BEJ2010}.
A careful comparison with other results found in the literature can be found in that reference.

We start by quoting our final results for the mass and the width of the $K^*(892)^{\pm}$
\bea
\label{massandwidth}
m_{K^*(892)^\pm} &=& 892.03\pm (0.19)_{\rm stat}\pm (0.44)_{\rm sys}\ \mbox{MeV}\ , \nn\\
\Gamma_{K^*(892)^\pm} &=& 46.53 \pm (0.38)_{\rm stat}\pm (1.0)_{\rm sys}\ \mbox{MeV}\ .
\eea
These results are obtained from the complex pole position on the
second Riemann sheet,  $s_{K^*}$, following the definition
$\sqrt{s_{K^*}}= m_{K^*} -(i/2)\Gamma_{K^*}$ \cite{Escribano:2002iv}.
It is important to stress that the mass and width thus obtained are rather different from the parameters that enter our description of the phase of $F_+(s)$.
When comparing results from different works one must always be sure that the same definition is used in all cases.
In Ref.~\cite{BEJ2010}, we showed that our results are compatible with others {\it provided} the pole position prescription is employed for all the analyses.

The final results for the parameters $\lambda_+'$ and $\lambda_+''$ read
\bea
\label{lambdas}
\lambda_+'\times 10^{3}  &=&   25.49\pm (0.30)_{\rm stat} \pm (0.06)_{s_{\rm cut}}\,, \nn\\
\lambda_+''\times 10^{4}  &=&  12.22\pm (0.10)_{\rm stat} \pm (0.10)_{s_{\rm cut}}\,.
\eea
In this case, the uncertainty from the variation of $s_{\rm cut}$ contributes as indicated.
From the expansion of Eq~(\ref{dispFF}) we can calculate the third coefficient of a Taylor series
of the type of Eq.~(\ref{FFTaylor}).
We find 
\beq
\lambda_+'''\times 10^5=8.87\pm (0.08)_{\rm stat} \pm (0.05)_{s_{\rm cut}}\,.
\eeq
These results are in good agreement with other analyses but have
smaller uncertainties since our fits are constrained by $\tauKpi$ and
$K_{l3}$ experiments.

In the extraction of $|V_{us}|$ from the $K_{l3}$ decay widths,
one must perform phase-space integrals where the form-factors play the central role.
The integrals are defined in Ref.~\cite{LR84,Antonelli10}.
From our form-factors we obtain the following results
\bea
I_{K^0_{e 3}}  =  0.15466(17)\ ,\quad I_{K^0_{\mu 3}}  =  0.10276(10)\ ,\nn\\
I_{K^+_{e 3}}  =  0.15903(17)\ ,\quad I_{K^+_{\mu 3}}  =  0.10575(11)\ . 
\eea
The uncertainties were calculated with a MC sample of parameters obeying the results of our fits with the correlations properly included.
The final uncertainties are competitive if compared with the averages of \cite{Antonelli10}
and the central values agree.

Another interesting result that can be extracted from the $\tauKpi$ spectrum is the $K\pi$ isospin-1/2 $P$-wave scattering phase.
The decay in question is indeed a very clean source of information about $K\pi$ interactions,
since the hadrons are isolated in the final state.
Below inelastic thresholds, the phase of the form-factor is the scattering phase,
as dictated by Watson's theorem.
From the expansion of the corresponding partial-wave $T$-matrix
element in the vicinity of the $K\pi$ threshold one can determine the
$K\pi$ $P$-wave threshold parameters. With our results, the first three
read
\bea
\label{ScattLengRes}
m_{\pi^-}^3 \, a_1^{1/2} \times 10		&=&	0.166(4)\ , \nn\\
m_{\pi^-}^5 \, b_1^{1/2} \times 10^{2}	&=&	0.258(9)\ , \nn\\
m_{\pi^-}^7\,  c_1^{1/2} \times 10^{3}	&=&  0.90(3)\ .   
\eea

\Acknowledgements
R.~E.~would like to express his gratitude to the CKM2010 Organizing Committee
for the opportunity of presenting this contribution,
and for the pleasant and interesting workshop we have enjoyed.
We are grateful to the Belle collaboration in particular to
S.~Eidelman, D.~Epifanov and B.~Shwartz, for providing their data.  
R.~E.~also acknowledges B.~Sciascia and F.~Mescia for useful discussions.


\begin{thebibliography}{99}


 \bibitem{LR84} H.~Leutwyler and M.~Roos,
  Z.\ Phys.\  C {\bf 25} (1984) 91.

\bibitem{JOP4} M.~Jamin, J.~A.~Oller and A.~Pich,
  Phys.\ Rev.\  D {\bf 74}, (2006) 074009 
  [arXiv:hep-ph/0605095].

 \bibitem{Belle}  D.~Epifanov {\it et al.}  [Belle Collaboration],
  Phys.\ Lett.\  B {\bf 654} (2007) 65 
  [arXiv:0706.2231 [hep-ex]].

\bibitem{Babar} 
S.~Paramesvaran [BaBar Collaboration], proceedings of Meeting of the
Division of Particles and Fields of the American Physical Society (DPF
2009), Detroit, Michigan, 26-31 Jul 2009, arXiv:0910.2884 [hep-ex].

\bibitem{BESIII}
  D.~M.~Asner {\it et al.}, ``Physics at BES-III'',
  arXiv:0809.1869 [hep-ex].
Edited by K.~T.~Chao and Y.~F.~Wang,
Int. J. of Mod. Phys. A {\bf 24} Supplement 1, (2009).

\bibitem{BEJ}  D.~R.~Boito, R.~Escribano and M.~Jamin,
  Eur.\ Phys.\ J.\  C {\bf 59}, (2009) 821
  [arXiv:0807.4883 [hep-ph]];
  PoS  {\bf EFT09}  (2009) 064
  [arXiv:0904.0425 [hep-ph]]

\bibitem{BEJ2010} D.~R.~Boito, R.~Escribano and M.~Jamin,
  JHEP {\bf 1009} (2010) 031
  [arXiv:1007.1858 [hep-ph]];
  Nucl.\ Phys.\ Proc.\ Suppl.\  {\bf 207-208} (2010) 148
  [arXiv:1010.0100 [hep-ph]];
  arXiv:1012.3493 [hep-ph].

\bibitem{Lattice}
  G.~Colangelo {\it et al.},
  arXiv:1011.4408 [hep-lat].

\bibitem{Omnes} R.~Omn\`es,
  Nuovo Cim.\  {\bf 8} (1958) 316.

\bibitem{PP2001}A.~Pich and J.~Portol\'es,
  Phys.\ Rev.\  D {\bf 63} (2001)  093005 
  [arXiv:hep-ph/0101194].

\bibitem{Bernardetal}
 V.~Bernard, M.~Oertel, E.~Passemar and J.~Stern,
  Phys.\ Lett.\  B {\bf 638} (2006)  480  
  [arXiv:hep-ph/0603202];
  Phys.\ Rev.\  D {\bf 80} (2009) 034034  
  [arXiv:0903.1654 [hep-ph]].

\bibitem{JPP20062008} M.~Jamin, A.~Pich and J.~Portol\'es,
  Phys.\ Lett.\  B {\bf 640} (2006) 176 
  [arXiv:hep-ph/0605096];
  Phys.\ Lett.\  B {\bf 664} (2008) 78 
  [arXiv:0803.1786 [hep-ph]].

\bibitem{JOP1} M.~Jamin, J.~A.~Oller and A.~Pich,
  Nucl.\ Phys.\  B {\bf 622} (2002) 279 
  [arXiv:hep-ph/0110193].

\bibitem{Antonelli10}
  M.~Antonelli {\it et al.},
  Eur.\ Phys.\ J.\  C {\bf 69} (2010) 399
  [arXiv:1005.2323 [hep-ph]].

\bibitem{Escribano:2002iv}
  R.~Escribano, A.~Gallegos, J.~L.~Lucio M, G.~Moreno and J.~Pestieau,
  Eur.\ Phys.\ J.\  C {\bf 28} (2003) 107
  [arXiv:hep-ph/0204338].

\end{thebibliography}
\end{document}